\documentclass[doublecol]{epl2}

\usepackage{amssymb,amsmath,latexsym}
\usepackage{accents}

\newcommand\al{\alpha}
\newcommand\be{\beta}
\newcommand\ga{\gamma}
\newcommand\de{\delta}
\newcommand\ep{\epsilon}

\newcommand\si{\sigma}
\newcommand\cc{{\cal M}}
\newcommand\tc{\tilde c}
\newcommand\tv{\tilde v}
\newcommand\tp{\tilde \phi}
\newcommand\tP{\tilde \Phi}
\newcommand\dg{\dagger}
\newcommand\pa{\partial}
\newcommand\beq{\begin{equation}}
\newcommand\eeq{\end{equation}}
\newcommand\bea{\begin{eqnarray}}
\newcommand\eea{\end{eqnarray}}
\newcommand\bi{\begin{itemize}}
\newcommand\ei{\end{itemize}}
\newcommand\ben{\begin{enumerate}}
\newcommand\een{\end{enumerate}}
\newcommand\non{\nonumber}
\newcommand\noi{\noindent}

\newcommand\bib{\bibitem}

\newcommand\ie{{\textit{i.e.}}}

\newcommand\rgd{{\textsf{RG}}}

\newif\ifboo \boofalse

\newcommand\ua{$\uparrow$}
\newcommand\da{$\downarrow$}
\newcommand\nufh{$\nu=5/2$~}

\newcommand\nuo{$\nu=1$~}
\newcommand\nuod{$\nu=1$}

\newcommand\nuoh{$\nu=1/2$~}
\newcommand\nuohd{$\nu=1/2$}
\newcommand\nuot{$\nu=1/3$~}

\newcommand\pfd{{\textsf{Pf}}}
\newcommand\pf{{\textsf{Pf~}}}

\title{Effect of inter-edge Coulomb interactions on transport through a point 
contact in a $\nu=5/2$ quantum Hall state}
\shorttitle{Inter-edge Coulomb interactions and $\nu=5/2$ quantum Hall state}

\author{Sourin Das\inst{1,\,2} \and Sumathi Rao\inst{3} \and Diptiman 
Sen\inst{4}}
\shortauthor{S. Das \etal}

\institute{
\inst{1} {Institut f\"ur Festk{\"o}rper-Forschung -- Theorie 3, 
Forschungszentrum J{\"u}lich, 52425 J{\"u}lich, Germany} \\
\inst{2} {Institut f\"ur Theoretische Physik A, RWTH Aachen, 52056 Aachen, 
Germany} \\
\inst{3} {Harish$-$Chandra Research Institute, Chhatnag Road, Jhusi, 
Allahabad 211 019, India} \\
\inst{4} {Centre for High Energy Physics, Indian Institute of Science,
Bangalore 560 012, India

E-mail: s.das@fz-juelich.de;~sumathi@hri.res.in;~diptiman@cts.iisc.ernet.in}}


\abstract{We study transport across a point contact separating two line 
junctions in a $\nu = 5/2$ quantum Hall system. We analyze the effect of 
inter-edge Coulomb interactions between the chiral bosonic edge modes of the 
half-filled Landau level (assuming a Pfaffian wave function for the 
half-filled state) and of the two fully filled Landau levels. In the presence 
of inter-edge Coulomb interactions between all the six edges participating in 
the line junction, we show that the stable fixed point corresponds to a point 
contact which is neither fully opaque nor fully transparent. Remarkably, this 
fixed point represents a situation where the half-filled level is fully 
transmitting, while the two filled levels are completely backscattered; hence 
the fixed point Hall conductance is given by $G_H=\frac{1}{2} e^2/h$. We 
predict the
non-universal temperature power laws by which the system approaches the stable
fixed point from the two unstable fixed points corresponding to the fully 
connected case ($G_H=\frac{5}{2} e^2/h$) and the fully disconnected case 
($G_H=0$).}

\begin{document}

\maketitle


In the last few years, there has been an upsurge in research on quantum Hall 
(QH) systems in the context of non-abelian QH states, primarily because they 
open up unprecedented possibilities for realizing robust topological quantum 
computers \cite{mic}. One of the most promising candidates for the 
experimental realization of a non-abelian QH state is the \nufh state which 
was initially believed to be described by a Pfaffian (\pfd) state \cite{moore},
but now has a few other states \cite{read} as plausible competitors.

Amongst the possible non-abelian QH states corresponding to various
filling fractions, the \nufh state is relatively easier to access in
an experiment. Hence, most theoretical proposals and experimental
attempts related to probing of the non-abelian QH states 
focus on the \nufh state, and they are primarily related to transport
measurements on the edge states. A complication that arises while making 
measurements on edge state corresponding to the half-filled
Landau level ($half$-$edge$), which hosts the non-abelian quasi-particle,
is that it is always masked by the two edge states corresponding to the two 
filled Landau levels. This is quite unlike the edge states of any fractional
QH state in the first Landau level, for example, the \nuot state. So
conductance measurements on such states will be in general difficult 
as far as directly probing them is concerned. 

In this Letter, we propose a set-up involving a point contact embedded in a 
line junction which naturally gets rid of the two filled Landau levels by 
fully backscattering them at the point contact, and lets the $half$-$edge$ 
transmit perfectly, without the need for fine-tuning of any gate voltage 
controlling transport across the point contact. When inter-edge interactions 
are introduced in the point contact geometry, the renormalization group flow 
automatically takes the system to this situation, which turns out to be the
stable fixed point of the interacting edge theory.

Another motivation for studying the point contact geometry with
inter-edge Coulomb interactions comes from the experiments of Dolev
\etal\cite{moty} and Radu \etal\cite{marcus}. The shot noise measurements by 
Dolev\etal~ confirmed the charge of the quasi-particle excitation in the 
half-filled Landau level in the $\nu=5/2$
state to be $e/4$. This measurement is unlikely to get influenced by the
presence of inter-edge Coulomb interaction. But, at the same time, this
measurement by itself does not confirm the non-abelian nature of the state.
On the other hand, the experiment by Radu \etal~ attempted to measure the
power law scaling exponents associated with quasi-particle tunneling
across a point contact. This measurement accompanied by the previous
one on the quasi-particle charge can definitely help to decide which of the
proposed candidate wave functions actually describe the $\nu=5/2$
state QH state. But the problem is that the power law scaling
exponents can get influenced by the inter-edge Coulomb
interactions, leading to non-universal values. Indeed, the experiment by
Radu \etal~ observed non-universal values for the power laws. One of
the prime reasons for such non-universal behavior could lie in the details of
the inter-edge Coulomb interaction. Hence, this aspect of the problem is
worth exploring. In general, the effects of inter-edge Coulomb interactions on
the transport in edge states across a point contact have been studied in the 
past and have been shown to be non-trivial \cite{pryadko,papa}.

We will now study the conductance across a point contact in the presence of 
inter-edge Coulomb interactions between both co-propagating and counter
propagating edge modes of the half-filled Landau level (assuming a
\pf wave function\cite{fisher}) and the two fully filled Landau
levels (assuming a chiral bosonic description for the edges), within a sharp 
edge scenario (\ie, with no edge reconstruction). Since a line 
junction is made by using metallic gates, the long-range part of the Coulomb
interaction is screened. We will therefore model the Coulomb potential by
a short-range density-density interaction. As mentioned above, for 
weak inter-edge Coulomb interactions, we will show that the stable fixed point
corresponds to an intermediate state in which the edge modes of the fully
filled Landau levels are completely backscattered, while the edge
corresponding to the half-filled Landau levels is completely transmitting
at the point contact. This results in a fixed point conductance
of $e^2/(2h)$ in the low temperature limit. This situation is to be contrasted
with the set-up where a point contact is
pinched off in a QH liquid with a single edge state, for example the
\nuot QH state \cite{kane1}. Here it is known that there are only
two fixed points (FPs), one corresponding to a perfectly transmitting point 
contact which is unstable, and the other corresponding to a completely
opaque point contact which is stable (for weak inter-edge
interactions). One can invert the stability of the two FPs by
introducing strong enough inter-edge interactions in a line junction
geometry \cite{kane1}, but no fixed point with intermediate
transmission emerges in the theory. But in the case of the \pf edge,
because of the multiple edge structure, there are fixed points which
are neither fully transmitting nor fully opaque. Once we introduce
even weak inter-edge Coulomb interaction, remarkably one such fixed
point with intermediate transmission and reflection becomes the stable one.

In our theoretical model each edge of a \nufh QH state consists of an
edge corresponding to \nuoh and, say, spin \ua, an edge with \nuo and
spin \da, and an edge with \nuo and spin \ua~ as we move outwards starting
from the bulk. [This spatial ordering of the edges follows from the decrease
in density on moving from the bulk towards the edge, and from the
Zeeman splitting due to the coupling of the electron spin to the
magnetic field.] The edge corresponding to \nuoh is described by a
charged chiral boson $\phi_1$ and a neutral (Majorana) fermion $\xi$
which has scaling dimension 1/2; in addition, an Ising spin field
$\si$ with scaling dimension 1/16 is required to describe the
quasi-particles with charge $e/4$ \cite{fisher}. The electron
annihilation operator on the \nuoh edge is given by $\psi_{e,1} =
\xi e^{i\sqrt{2} \phi_1}$ and has scaling dimension $1/2 + 1 = 3/2$
in the absence of interactions. The quasi-particle annihilation
operator on the \nuoh edge is given by $\psi_{qp,1} = \si
e^{i\phi_1/ (2\sqrt{2})}$, and has scaling dimension $1/16 + 1/16 =
1/8$ in the absence of interactions. The two edges with \nuo can be
written in terms of charged chiral bosons $\phi_2$ and $\phi_3$. The
electron operators on these edges are given by $\psi_{e,a} =
e^{i\phi_a}~$ ($a=2,3$), and both their scaling dimensions are $1/2$
in the absence of interactions. Since the Majorana fermion $\xi$ and
the Ising spin $\si$ are neutral, their scaling dimensions will not
change when we introduce Coulomb interactions between the charged fields. 
However, the scaling dimensions of operators like $e^{iq_a \phi_a}$ depend 
on the strength of the interactions between counter propagating fields and 
will change when Coulomb interactions are introduced.


\begin{figure}[htb]
\includegraphics[height=3.4in,width=3.4in]{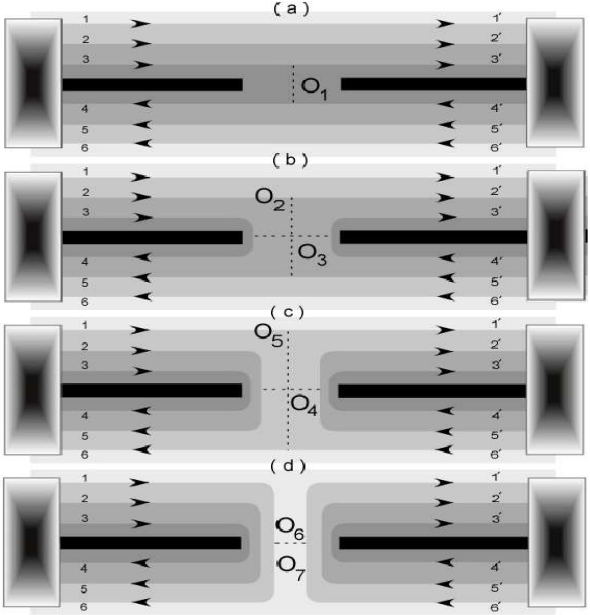}
\caption{Pictures of different configurations of two \nufh edges
split by line junctions which are themselves separated by a point
contact in the middle ($x=0$). The six edges (three incoming and
three outgoing) are marked as $1 - 6$ on the left and $1' - 6'$ on
the right; they are connected to each other at $x=0$ in four possible ways 
shown in (a-d). The black regions denote the line junctions; the other shades 
show the filling fractions 0, 1, 2 and 5/2 from darkest to lightest, the 
white regions denoting the $\nu = 5/2$ bulk.} \end{figure}

We now consider what happens if a \nufh QH liquid is split into two
by applying a gate voltage along two semi-infinite lines with a gap
(point contact) separating them; this depletes the electron density
along those regions which are called line junctions. We denote the
six edges from top to bottom of a \nufh edge with a line junction as
1 (\nuoh edge, spin \ua), 2 (\nuod, spin \da), 3 (\nuod, spin \ua),
4 (\nuod, spin \ua), 5 (\nuod, spin \da), and 6 (\nuohd, spin \ua);
the edges of the other \nufh edge with a line junction will be
similarly denoted as $1', 2', \cdots, 6'$. The different edges of
the two line junctions can now be connected to each other in several
ways. Assuming that only edges with the same spin can connect to
each other and that there are no crossings between different edges,
there are four possible configurations as shown in Figs. 1 (a-d);
all of them represent fixed points in the theory. They correspond to
the following boundary conditions at the location of the point
contact, called $x=0$, between the different bosonic fields: (a)
$\phi_a = \phi_{a'}$ for $a=1,2,\cdots,6$, (b) $\phi_a = \phi_{a'}$
for $a=1,2,5,6$, $~\phi_3 = \phi_4$, and $\phi_{3'} = \phi_{4'}$,
(c) $\phi_1 = \phi_{1'}$, $\phi_6 = \phi_{6'}$, $\phi_2 = \phi_5$,
$\phi_3 = \phi_4$, $\phi_{2'} = \phi_{5'}$, and $\phi_{3'} =
\phi_{4'}$, and (d) $\phi_1 = \phi_6$, $\phi_2 = \phi_5$, $\phi_3 =
\phi_4$, $\phi_{1'} = \phi_{6'}$, $\phi_{2'} = \phi_{5'}$, and
$\phi_{3'} = \phi_{4'}$. These conditions can be written in terms of
a current splitting matrix $\cc$ as follows. We define two vectors
$\phi_I$ and $\phi_O$, denoting incoming and outgoing fields, as
$~\phi_{iI} = (\phi_1 ~\phi_2 ~\phi_3 ~\phi_{6'} ~\phi_{5'}
~\phi_{4'})$, and $~\phi_{iO} = (\phi_4 ~\phi_5 ~\phi_6 ~\phi_{3'}
~\phi_{2'} ~\phi_{1'})$. These are related at $x=0$ by a $6 \times
6$ real orthogonal matrix $\cc$ as $\phi_{iO} = \sum_j \cc_{ij} \phi_{jI}$.
Current conservation implies that each column of $\cc$ must add up to 1.






We will study the stability of the configurations in Fig. 1 with respect to 
tunneling between different edges. To do this, we will consider the following 
tunneling operators at the point $x=0$ between nearby edges with the same 
spin in the configurations shown in Figs. 1 (a-d):

\noi (a) $O_1 = \psi_{e,4}^\dg \psi_{e,3} = \exp [i (\phi_3 - \phi_4)]$,

\noi (b) $O_2 = \psi_{e,5}^\dg \psi_{e,2} = \exp [i (\phi_2 - \phi_5)]$, and 
$O_3 = \psi_{e,3'}^\dg \psi_{e,3} = \exp [i (\phi_3 - \phi_{3'})]$,

\noi (c) $O_4 = \psi_{e,2'}^\dg \psi_{e,2} = \exp [i (\phi_2 -
\phi_{2'})]$, and $O_5 = \psi_{e,6}^\dg \psi_{e,1} = \xi_1 \xi_6
\exp [i \sqrt{2} (\phi_1 - \phi_6)]$,

\noi (d) $O_6 = \psi_{e,1'}^\dg \psi_{e,1} = \xi_1 \xi_{1'} \exp [i
\sqrt{2} (\phi_1 - \phi_{1'})]$, and $O_7 = \psi_{qp,1'}^\dg \psi_{qp,1} 
= \si_1 \si_{1'} \exp [i (\phi_1 - \phi_{1'})/(2\sqrt{2})]$.

\noi Of these seven operators, $O_1 - O_6$ correspond to electron
tunneling, while $O_7$ corresponds to quasi-particle tunneling. In
the absence of interactions, we can easily compute the scaling
dimensions $d_i$ of the operators $O_i$. We find that $d_1 = d_2 =
d_3 = d_4 = 1$ (marginal), $d_5 = d_6 = 3$ (irrelevant), and $d_7 =
1/4$ (relevant). For weak interactions, $O_5$ and $O_6$ ($O_7$) will
continue to remain irrelevant (relevant); thus Fig. 1 (d) is an
unstable FP. We will study below whether the operators $O_1 - O_4$
will become relevant or irrelevant for weak interactions. We note that if 
there are density-density interactions only between co-propagating modes, 
then none of the scaling dimensions would get modified. It is therefore 
important to consider interactions between counter propagating modes as well.


In any of the configurations shown in Fig. 1, there are three right
moving and three left moving bosonic fields either far to the left
or far to the right of the point contact. Ignoring the two Majorana
fermion fields, the Lagrangian density of the six bosonic fields can
be written as \beq {\cal L} = - \frac{1}{4\pi} ~\left( \sum_{a=1}^6
\ep_a \pa_t \phi_a \pa_x \phi_a + \sum_{a,b=1}^6 \pa_x \phi_a K_{ab}
\pa_x \phi_b \right)~, \label{lag2} \eeq where $\ep_a = 1$ ($-1$)
for the three right (left) moving fields respectively. In Eq.
(\ref{lag2}), we have absorbed the velocities in the diagonal
elements $K_{aa}$ for simplicity; three of the velocities will be
positive and three negative. [For repulsive interactions between
edges $a$ and $b$, we have $\ep_a \ep_b K_{ab} > 0$]. The densities
and currents are given by $\rho_a = (\ep_a /2\pi) \pa_x \phi_a$ and $j_a 
= - (\ep_a /2\pi) \pa_t \phi_a$; these satisfy the equations of continuity.
To quantize the theory, we impose the equal-time commutation relations 
$[\phi_a (x) , \rho_b (y)] = -i \de_{ab} \de (x-y)$. These will be satisfied 
if the fields are given at time $t=0$ by $\phi_a (x) = \int_0^\infty 
\frac{dk}{k}$ $[c_{ak} e^{i \ep_a kx} + c_{ak}^\dg e^{-i \ep_a kx}]$, where 
$[c_{ak}, c_{bk'}^\dg ] = \de_{ab} ~k ~\de (k-k')$.

The Lagrangian in Eq. (\ref{lag2}) can be diagonalized either by a
Bogoliubov transformation \cite{tsallis} or, equivalently, by
solving the equations of motion. We assume that the fields take the
form $\phi_a = X_{a\al} e^{ik(x - \tv_\al t)}$, where the index
$\al$ ($= 1,2,\cdots,6$) labels the different solutions, and
$\tv_\al$ are the corresponding velocities. The $X_{a\al}$ (which
are real) and the $\tv_\al$ can be obtained by solving the equations
\beq \sum_{b=1}^6 ~\ep_a ~K_{ab} ~X_{b\al} ~=~ \tv_\al ~X_{a\al}.
\label{eigen} \eeq If the problem is well defined (namely, the
interactions between the counter propagating modes are not too
strong compared to the interactions between co-propagating modes),
the velocities $\tv_\al$ will all be real, with three of them being
positive (right moving) and three being negative (left moving). We
introduce a label $\ep_\al = 1$ and $-1$ for the eigenmodes with $\tv_\al
> 0$ and $< 0$ respectively. The eigenvectors can then be normalized so that
\bea \sum_{a=1}^6 ~\ep_a \ep_\al ~X_{a\al} ~X_{a\be} &=& \de_{\al\be}~, \non \\
\sum_{\al=1}^6 ~\ep_a \ep_\al ~X_{a\al} ~X_{b\al} &=& \de_{ab}~.
\label{norm} \eea 
We introduce the projection operators $P_{a\al,\pm} = (1 \pm \ep_a \ep_\al)/2$.
The original and new (Bogoliubov transformed) annihilation operators $c_{ak}$ 
and $\tc_{\al k}$ are related as 
\bea \tc_{\al k} &=& \sum_{a=1}^6 ~X_{a\al} ~[ P_{a\al,+} ~c_{ak} ~-~ 
P_{a\al,-} ~c_{ak}^\dg ]~, \non \\
c_{ak} &=& \sum_{\al=1}^6 ~X_{a\al} ~[ P_{a\al,+} ~\tc_{\al k} ~+~ P_{a\al,-} ~
\tc_{\al k}^\dg ]~. \label{transf} \eea 
Using Eq. (\ref{norm}), we can verify that $[c_{ak}, c_{bk'}] = 0$ and
$[c_{ak}, c_{bk'}^\dg ] = \de_{ab} ~k ~\de (k-k')$ imply that
$[\tc_{\al k}, \tc_{\be k'}] =0$ and $[\tc_{\al k}, \tc_{\be k'}^\dg
] = \de_{\al \be} ~k~ \de (k-k')$. Eq. (\ref{transf}) implies that
the original and new bosonic fields are related at the point $x=0$
as $~\tp_\al = \sum_{a=1}^6 \ep_a \ep_\al X_{a\al} \phi_a$ and
$~\phi_a = \sum_{\al=1}^6 X_{a\al} \tp_\al$.

We will choose the same values of the parameters $K_{ab}$ for the
six edges on the left and on the right of the configurations in Fig.
1; this ensures that the same matrix $X_{a\al}$ will govern the
fields on both sides of the point contact, $\phi_a$, $\tp_\al$ and
$\phi_{a'}$, $\tp_{\al'}$. Further, we will assume for simplicity
that $K_{ab}$ is mirror symmetric under a reflection about the line
junction, \ie, $K_{a,b} = K_{7-a,7-b}$ for all values of $a$ and
$b$. For any such choice of the parameters $K_{ab}$, we can compute
the scaling dimensions of the tunneling operators $O_1 - O_7$
mentioned earlier as follows. We first find the eigenvectors
$X_{a\al}$ as indicated in Eqs. (\ref{eigen}) and (\ref{norm}). We
can then relate the Bogoliubov transformed incoming and outgoing
fields to the original incoming and outgoing fields. These take the forms 
\bea \tp_{iI} &=& \sum_{j=1}^6 ~\left[ A_{ij} ~\phi_{jI} ~-~
B_{ij} ~\phi_{jO} \right]~, \non \\
\tp_{iO} &=& \sum_{j=1}^6 ~\left[ C_{ij} ~\phi_{jO} ~-~ D_{ij}
~\phi_{jI} \right]~, \label{abcd} \eea 
where the matrices $A ~-~ D$ can be constructed as follows. If the $6 \times 
6$ matrix $X_{a\al}$ has a block form
\beq X ~=~ \left( \begin{array}{cc} X_1 & X_2 \\
X_3 & X_4 \end{array} \right)~, \eeq where the $X_i$'s are $3 \times
3$ matrices, the matrices $A - D$ have the block forms
\bea A &=& \left( \begin{array}{cc} X_1 & 0 \\
0 & X_1 \end{array} \right) , ~~~~B ~=~ \left( \begin{array}{cc} X_2 & 0 \\
0 & X_2 \end{array} \right) , \non \\
C &=& \left( \begin{array}{cc} X_4 & 0 \\
0 & X_4 \end{array} \right) , ~~~~D ~=~ \left( \begin{array}{cc} X_3 & 0 \\
0 & X_3 \end{array} \right) . \eea Since $\phi_{iO}$ and $\phi_{iI}$
are related at $x=0$ as $\phi_O = \cc \phi_I$, Eq. (\ref{abcd})
implies that $\phi_I = (A - B\cc)^{-1} \tp_I$ and $\phi_O = \cc (A -
B\cc)^{-1} \tp_I$. This gives a relation between the twelve fields
$\phi_a$ and $\phi_{a'}$ and the six incoming Bogoliubov fields
$\tp_{iI}$ which can be taken to be mutually independent. Since the
scaling dimension of $\exp (i \sum_i q_i \tp_{iI})$ is $(1/2) \sum_i
q_i^2$, we can find the scaling dimension of the exponential of any
linear combination of the fields $\phi_a$ and $\phi_{a'}$.

The scaling dimensions of the operators $O_1 - O_7$ can now be
computed. The computation is somewhat involved for the
configurations shown in Figs. 1 (b) and 1 (c), but is simple for
Figs. 1 (a) and 1 (d). The scaling dimension of the operator $O_1 =
\exp [i (\phi_3 - \phi_4)]$ in Fig. 1 (a) is given by $d_1 = (1/2)
\sum_{a=1}^6 (X_{3a} - X_{4a})^2$, while the scaling dimensions of
the operators $O_6$ and $O_7$ in Fig. 1 (d) are given by $d_6 = 1 +
2 \ga$ and $d_7 = (1 + \ga)/8$ respectively, where $\ga = (X_{11} +
X_{16})^2 + (X_{12} + X_{15})^2 + (X_{13} + X_{14})^2$.

To be explicit, let us discuss a specific form of the interaction
parameters given by the matrix $K_{ab}$. We take all the diagonal entries
$K_{aa}$ equal to $v_0$, the off-diagonal entries corresponding to all pairs
of co-propagating modes equal to $v_1$, and the off-diagonal entries 
corresponding to all pairs of counter propagating modes equal to $-v_2$, 
where the three parameters $v_i$ are all positive (corresponding to repulsive 
interactions). Namely, $K$ has the highly symmetric form
\beq K_{ab} ~=~ \left( \begin{array}{cccccc}
v_0 & v_1 & v_1 & -v_2 & -v_2 & -v_2 \\
v_1 & v_0 & v_1 & -v_2 & -v_2 & -v_2 \\
v_1 & v_1 & v_0 & -v_2 & -v_2 & -v_2 \\
-v_2 & -v_2 & -v_2 & v_0 & v_1 & v_1 \\
-v_2 & -v_2 & -v_2 & v_1 & v_0 & v_1 \\
-v_2 & -v_2 & -v_2 & v_1 & v_1 & v_0 \end{array} \right). \eeq
We then find that the Bogoliubov transformation simplifies if
we first define the linear combinations
\bea \Phi_1 &=& \frac{1}{\sqrt 3} ~(\phi_1 ~+~ \phi_2 ~+~ \phi_3), \non \\
\Phi_2 &=& \frac{1}{\sqrt 2} ~(\phi_1 ~-~ \phi_2), \non \\
\Phi_3 &=& \frac{1}{\sqrt 6} ~(\phi_1 ~+~ \phi_2 ~-~ 2 ~\phi_3), \non \\
\Phi_4 &=& \frac{1}{\sqrt 3} ~(\phi_6 ~+~ \phi_5 ~+~ \phi_4), \non \\
\Phi_5 &=& \frac{1}{\sqrt 2} ~(\phi_6 ~-~ \phi_5), \non \\
\Phi_6 &=& \frac{1}{\sqrt 6} ~(\phi_6 ~+~ \phi_5 ~-~ 2 ~\phi_4). \eea
The corresponding Bogoliubov transformed fields are $\tP_i$. We find that 
$\tP_1$ and $\tP_4$ both have the velocity $\tv_1 = \sqrt{(v_0 + 2 v_1)^2 - 
9 v_2^2}$, and they are related to the original fields $\Phi_1$ and $\Phi_4$ as
\bea \tP_1 &=& \cosh \theta ~\Phi_1 ~-~ \sinh \theta ~\Phi_4, \non \\
\tP_4 &=& \cosh \theta ~\Phi_4 ~-~ \sinh \theta ~\Phi_1, \non \\
{\rm where}~~~ e^{-2\theta} &=& \sqrt{\frac{v_0 + 2 v_1 - 3 v_2}{v_0 + 
2 v_1 + 3 v_2}}. \eea
Note that $\theta > 0$ if the interaction $v_2$ between counter propagating
modes is repulsive. All the remaining fields $\Phi_{2,3,5,6}$ have the 
velocity $\tv_2 = v_0 - v_1$, and they are not affected by the Bogoliubov 
transformation. The validity of all these relations requires that $v_0 > v_1$
and $v_0 + 2 v_1 > 3 v_2$. We then find that the scaling dimensions of the
operators $O_1$, $O_6$ and $O_7$ are given by $d_1 = (2 + e^{-2\theta})/3$,
$d_6 = (7 + 2 e^{2\theta})/3$, and $d_7 = (5 + e^{2\theta})/24$. We see that
$O_1$ is relevant if $\theta > 0$, $O_6$ is always irrelevant, and $O_7$
is relevant as long as $e^{2\theta} < 19$, i.e., if the interaction $v_2$
is not very strong. The scaling dimensions of the other four operators
$O_{2,3,4,5}$ can be found numerically as indicated above. We find that
$O_2$ is relevant while $O_{3,4,5}$ are irrelevant for all values of $\theta
> 0$. This implies that the configuration in Fig. 1 (a) is unstable, 1 (d) 
is unstable if $v_2$ is not very large, 1 (b) is unstable, and 1 (c) is stable.
In Table 1, we present the scaling dimensions of the seven operators 
for a number of values of $e^{-2\theta}$ which is a measure of the
strength of the interactions between the counter propagating modes.
\vspace*{.2cm}

\begin{center} \begin{tabular}{|c|c|c|c|c|c|} \hline
$e^{-2\theta} \to$ & 0.2 & 0.4 & 0.6 & 0.8 & 1 \\ \hline
$d_1$ & 0.73 & 0.80 & 0.87 & 0.93 & 1 \\ \hline
$d_2$ & 0.64 & 0.75 & 0.85 & 0.93 & 1 \\ \hline
$d_3$ & 1.36 & 1.25 & 1.15 & 1.07 & 1 \\ \hline
$d_4$ & 1.57 & 1.33 & 1.18 & 1.08 & 1 \\ \hline
$d_5$ & 1.86 & 2.33 & 2.64 & 2.85 & 3 \\ \hline
$d_6$ & 5.67 & 4.00 & 3.44 & 3.17 & 3 \\ \hline
$d_7$ & 0.42 & 0.31 & 0.28 & 0.26 & 0.25 \\ \hline
\end{tabular} \end{center}

\noi Table 1. Scaling dimensions $d_i$ of the seven operators for a number of
values of $e^{-2\theta} \le 1$ (repulsive interactions).
\vspace*{.3cm}

We have numerically studied the effects of some other forms of the matrix 
$K_{ab}$. For a large range of interactions between all the edges, we find
that the configurations in Figs. 1 (a), (b) and (d) are
unstable to the tunnelings described by the operators $O_1$, $O_2$
and $O_7$, namely, their scaling dimensions $d_1$, $d_2$ and $d_7$
are all less than 1. (The operators $O_3$ and $O_6$ are irrelevant).
However, the configuration in Fig. 1 (c) is stable under the
tunnelings described by both $O_4$ and $O_5$, namely, their scaling
dimensions $d_4$ and $d_5$ are larger than 1. These results may be
expected since it is known that weak interactions make
backscattering between two perfectly transmitting \nuo edges
relevant, and tunneling between two perfectly reflecting \nuo edges
irrelevant \cite{kane2}. The stability of the configuration in Fig.
1 (c) has a remarkable consequence. If we measure the Hall
conductance $G$ between the left and right halves of the system,
only the \nuoh mode would contribute since the \nuo modes are
completely backscattered; we should therefore obtain a low-temperature value 
of $G$ equal to $e^2/(2h)$. An experimental confirmation of this would 
provide the first example in a QH system of such a value of $G$.

\begin{figure}[htb]
\hspace*{1cm}
\includegraphics[height=2.5in,width=2.5in]{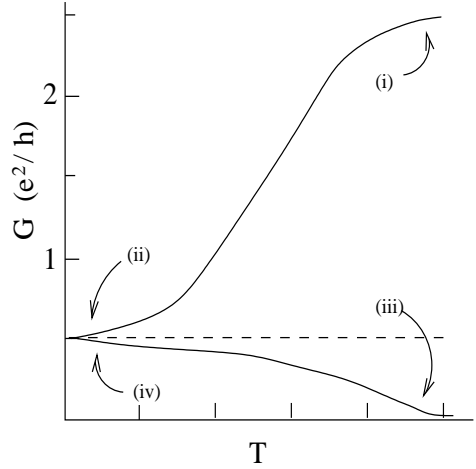}
\caption{Variation of Hall conductance $G$ (in units of $e^2/h$) with 
temperature $T$ (in arbitrary units) for two possible starting points at high 
temperature. The upper curve shows a RG flow from an unstable FP at 
$5e^2/(2h)$ to a stable FP at $e^2/(2h)$ at low temperature, while the lower 
curve shows a flow from an unstable FP at 0 to $e^2/(2h)$ at low temperature. 
The regions (i-iv) are explained in the text.} \end{figure}

Experimentally, two different possibilities exist depending on whether we 
start at high temperature with the configuration in Fig. 1 (a) (all modes 
fully transmitting) or 1 (d) (all modes fully backscattered). If we start
with Fig. 1 (a), we have $G = 5e^2 /(2h)$ at high temperature. Then
if switch on a small amount of backscattering at the point contact, then a
\rgd~ argument implies that as the temperature $T$ is decreased, $G$
will decrease from $5e^2 /(2h)$ as $1/T^{1-d_1}$ (region (i) in Fig.
2). This will continue till $G$ flows at low temperature to $e^2/(2h)$ as in 
Fig. 1 (c). Also note that, at some intermediate temperature, the
system will go through the configuration in Fig. 1 (b) where $G =
3e^2 /(2h)$.] As $T \to 0$, $G$ will approach $e^2/(2h)$ from above
as $T^{d_4 -1}$ (region (ii) in Fig. 2), since this results from the
vanishing of the irrelevant tunneling $O_4$. On the other hand, if
we start with Fig. 1 (d) at high temperature, $G$ is equal to 0 and
then switch on a small amount of tunneling at the point contact; it
will then increase as $1/T^{1-d_7}$ (region (iii) in Fig. 2) as the
temperature is reduced. Finally, it will approach $e^2/(2h)$ from
below as $T^{d_5 -1}$ at low temperature (region (iv) in Fig. 2),
since this follows from the vanishing of the irrelevant tunneling
$O_5$. From an experimental point of view, it might be expected that
the line junction formed by the gates would allow electrons to
tunnel across it, leading to a disorder dominated phase. But it may
be possible to avoid this by fine etching of the two-dimensional
electron gas along the line junction beneath the gates. This should
allow the scenario we have proposed to be experimentally accessible.
Note that we have not considered the outer edges of the $\nu = 5/2$
bulk which lie far away from the line junction as is depicted in 
Fig. 3. This would be justified in a transport measurement set-up in
which the top, bottom and the left contacts are at the same voltage, 
and the bias is applied at the right contact (see Fig. 3). This 
will ensure that the current that is collected at the left contact is 
solely due to the transport taking place across the line junction 
and not due to the outer two edges, and hence can confirm our prediction.

\begin{figure}[htb]
\includegraphics[height=2.25in,width=3.2in]{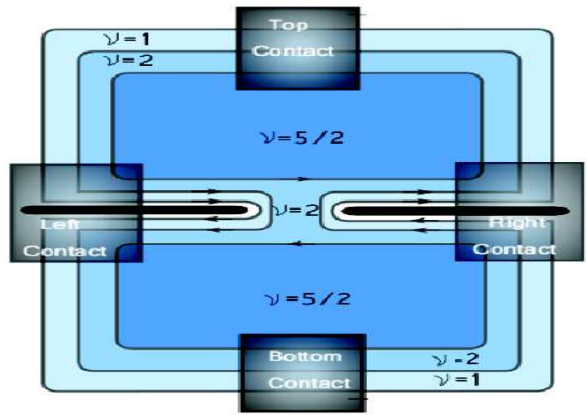}
\caption{Schematic picture of a four-terminal set-up which allows for the 
measurement of a conductance of $\frac{e^2}{2h}$ arising solely due to 
transport across the line junction alone and not the outer two top and bottom,
left and right moving edge states. The four gray patches represent the ohmic 
contacts. The deep blue region represents the $\nu=5/2$ Hall states, and the 
lighter shades represents the region of the other two integer filling 
fractions.} \end{figure}

To conclude, we have shown that for the \nufh state, the presence of
inter-edge Coulomb interactions can lead to a stable fixed point
with a fixed point conductance of $G_H=e^2/(2h)$, which uniquely
identifies a Hall plateau with an effective filling fraction of
half. We have predicted the power laws of the deviations of the 
conductance from its fixed point value, as functions of the interaction 
strengths between the various edge states. We have also provided 
an explicit experimental set-up which can check the predictions that we make 
in this letter. In general, the strategy outlined here can be applied to 
unmask any fractional Hall edge masked by integer Hall edge states. 

SD thanks Moty Heiblum for discussions. SD and DS thank DST, India for
financial support under Project No. SR/S2/CMP-27/2006.


\end{document}